\documentclass[epj]{svjour}
\usepackage{amsfonts}
\usepackage{amsmath}
\usepackage{graphicx}
\usepackage{dcolumn}
\usepackage{bm}
\usepackage{booktabs}
\usepackage{feynmf}   
\usepackage{epstopdf}   
\usepackage{slashed}  
\usepackage{cases}
\usepackage{multirow}
\usepackage{hyperref}
\usepackage{subfigure}
\usepackage{widetext}

\newcommand{\feynp}[1]{#1\kern-0.45em/}
\def\dab{\int^{\alpha_{max}}_{\alpha_{min}}d\alpha\int^{\beta_{max}}_{\beta_{min}}d\beta}
\def\qq{\langle\bar qq\rangle}
\def\sss{\langle\bar ss\rangle}
\def\GGa{\langle GG\rangle}
\def\GGb{\langle g_s^2GG\rangle}

\def\qGqa{\langle\bar qGq\rangle}

\def\qGqb{\langle g_s\bar q\sigma Gq\rangle}
\def\sGsb{\langle g_s\bar s\sigma Gs\rangle}
\def\FF(s){\left[(\alpha+\beta)m_c^2-\alpha\beta s\right]}
\def\HH(s){\left[m_c^2-\alpha(1-\alpha) s\right]}
\def\KK(s){\left[\gamma m_c^2-\gamma(1-\gamma) s\right]}
\def\non{\\ \nonumber}

\allowdisplaybreaks[3]

\begin{document}

\title{Establishing the first hidden-charm pentaquark with strangeness}

\author{Hua-Xing Chen\inst{1}\thanks{hxchen@buaa.edu.cn} \and Wei Chen\inst{2}\thanks{chenwei29@mail.sysu.edu.cn} \and Xiang Liu\inst{3,4}\thanks{xiangliu@lzu.edu.cn} \and Xiao-Hai~Liu\inst{5}\thanks{xiaohai.liu@tju.edu.cn}
}                     
\offprints{}          
\institute{
School of Physics, Southeast University, Nanjing 210094, China
\and
School of Physics, Sun Yat-Sen University, Guangzhou 510275, China
\and
School of Physical Science and Technology, Lanzhou University, Lanzhou 730000, China
\and
Research Center for Hadron and CSR Physics, Lanzhou University and Institute of Modern Physics of CAS, Lanzhou 730000, China
\and
Center for Joint Quantum Studies and Department of Physics, School of Science, Tianjin University, Tianjin 300350, China
}

\date{Received: date / Revised version: date}
%
\abstract{
We study the $P_{cs}(4459)^0$ recently observed by LHCb using the method of QCD sum rules. Our results support its interpretation as the $\bar D^* \Xi_c$ hadronic molecular state of either $J^P=1/2^-$ or $3/2^-$. Within the hadronic molecular picture, the three LHCb experiments observing $P_c$ and $P_{cs}$ states~\cite{lhcb,Aaij:2015tga,Aaij:2019vzc} can be well understood as a whole. This strongly supports the existence of hadronic molecules, whose studies can significantly improve our understanding on the construction of the subatomic world. To verify this picture, we propose to further investigate the $P_{cs}(4459)^0$ to examine whether it can be separated into two states, and to search for the $\bar D \Xi_c$ molecular state of $J^P=1/2^-$.
\PACS{
      {12.39.Mk}{Glueball and nonstandard multi-quark/gluon states} \and
      {12.38.Lg}{Other nonperturbative calculations}
     } 
} 
\maketitle

\begin{figure}[hbt]
\begin{center}
\scalebox{0.48}{\includegraphics{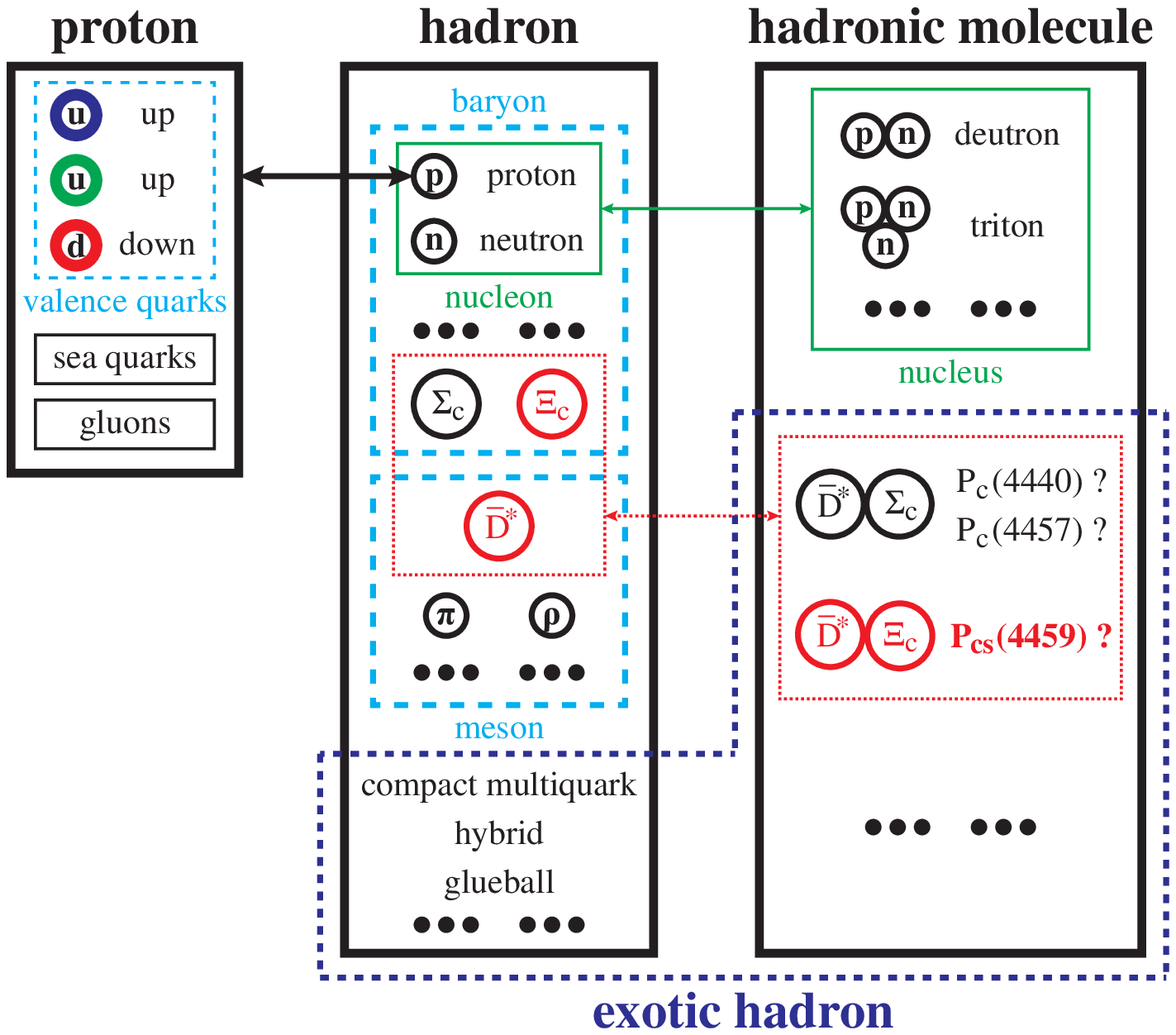}}
\caption{A possible way to construct the subatomic world.}
\label{fig:construction}
\end{center}
\end{figure}

$\\$
{\it Introduction.}-----
Atomic nuclei are made of protons and neutrons, which are themselves composed of quarks and gluons. In the past century a huge number of subatomic particles, called hadrons, were discovered in particle experiments, whose properties are similar to the proton and neutron~\cite{pdg}. One naturally raises an interesting question: are there subatomic particles corresponding to the nucleus? Nowadays we call them ``hadronic molecules'', whose studies can significantly improve our understanding on the construction of the subatomic world, as illustrated in Fig.~\ref{fig:construction}.

It is not so easy to answer the above question. Most of the experimentally observed hadrons can be described as $q\bar q$ mesons or $qqq$ baryons in the conventional quark model, while there can also exist (compact) $q\bar qq\bar q$ tetraquarks and $qqqq\bar q$ pentaquarks~\cite{GellMann:1964nj,Zweig:1981pd}, etc. With the experimental progress on this issue over the past decade, dozens of $XYZ$ charmonium-like states were reported, providing us good opportunities to identify exotic hidden-charm tetraquarks~\cite{pdg}. However, these states may be explained as hadronic molecules, while they may also be explained as compact tetraquarks that are still hadrons.

Fortunately, in the LHCb experiments performed in 2015 and 2019~\cite{Aaij:2015tga,Aaij:2019vzc}, the famous hidden-charm pentaquark states, $P_c(4312)$, $P_c(4380)$, $P_c(4440)$, and $P_c(4457)$, were discovered. These four $P_c$ states contain at least five quarks $\bar c c uud$, so they are perfect candidates of hidden-charm pentaquark states. Among them, the three narrow states $P_c(4312)$, $P_c(4440)$, and $P_c(4457)$ are just below the $\bar D \Sigma_c$ and $\bar D^{*} \Sigma_c$ thresholds, so their natural interpretations are the $\bar D \Sigma_c$ and $\bar D^{*} \Sigma_c$ hadronic molecular states~\cite{Chen:2015loa,Chen:2015moa,Roca:2015dva,Meissner:2015mza,Liu:2019tjn}, whose existence had been predicted in Refs.~\cite{Wu:2010jy,Wang:2011rga,Yang:2011wz,Li:2014gra,Karliner:2015ina} before the LHCb experiment performed in 2015~\cite{Aaij:2015tga}. However, there still exist other possible explanations~\cite{Maiani:2015vwa,Lebed:2015tna,Wang:2015epa,Guo:2015umn,Liu:2015fea,Bayar:2016ftu}, and we refer to the reviews~\cite{Chen:2016qju,Liu:2019zoy,Guo:2017jvc,Ali:2017jda,Olsen:2017bmm,Brambilla:2019esw} for detailed discussions.

It is natural to conjecture whether the hidden-charm pentaquark state with strangeness exists or not. Such a state is usually denoted as ``$P_{cs}$'', whose quark content is $\bar c c s q q$ ($q=u/d$). There have been some but not many theoretical studies on it~\cite{Cheng:2015cca,Anisovich:2015zqa,Wang:2015wsa,Feijoo:2015kts,Lu:2016roh,Chen:2016ryt,Xiao:2019gjd,Zhang:2020cdi,Shen:2020gpw}. Especially, we proposed in Ref.~\cite{Chen:2015sxa} to search for the $P_{cs}$ in the $J/\psi \Lambda$ invariant mass spectrum of the $\Xi_b^- \to J/\psi K^- \Lambda$ decays. Besides, in Ref.~\cite{Wang:2019nvm} the authors studied the $P_{cs}$ using the chiral effective field theory, and calculated the mass of the $J^P = 1/2^-$ $\bar D^{*} \Xi_c$ molecular state to be $4456.9^{+3.2}_{-3.3}$~MeV. Note that both of these two references are based on the hadronic molecular picture.

Very recently, the LHCb Collaboration reported the evidence of a hidden-charm pentaquark state with strangeness, $P_{cs}(4459)^0$, in the $J/\psi \Lambda$ invariant mass spectrum of the $\Xi_b^- \to J/\psi K^- \Lambda$ decays~\cite{lhcb}. Its mass and width were measured to be:
\begin{eqnarray}
P_{cs}(4459)^0:    &M=& 4458.8 \pm  2.9^{+4.7}_{-1.1}  \mbox{ MeV} \, ,
\label{experiment}
\\ \nonumber                &\Gamma=& 17.3 \pm  6.5^{+8.0}_{-5.7}    \mbox{ MeV} \, ,
\end{eqnarray}
while its spin-parity quantum number was not determined since the statistic	is not enough. Note that the channel observing the $P_{cs}(4459)^0$ is just the one proposed by us in Ref.~\cite{Chen:2015sxa}, and the above mass value is almost identical to the mass of the $J^P = 1/2^-$ $\bar D^{*} \Xi_c$ molecular state predicted in Ref.~\cite{Wang:2019nvm}. Therefore, the present LHCb experiment~\cite{lhcb} strongly supports the hadronic molecular picture once more.

Actually, as indicated by LHCb, the $P_{cs}(4459)^0$ is about 19~MeV below the $\bar D^{*0} \Xi_c^0$ threshold~\cite{lhcb}, so it is natural to interpret it as the $\bar D^{*} \Xi_c$ molecular state, with the spin-parity quantum number $J^P = 1/2^-$ or $3/2^-$. Accordingly, in this letter we shall study the $\bar D^{*} \Xi_c$ molecular states of $J^P = 1/2^-$ and $3/2^-$ using the method of QCD sum rules, and at the same time we shall also study the $\bar D \Xi_c$ molecular state of $J^P = 1/2^-$. We calculate masses of $\bar D^* \Xi_c$ molecular states to be $4.46^{+0.16}_{-0.14}$~GeV for the $J^P = 1/2^-$ one and $4.47^{+0.19}_{-0.15}$~GeV for the $J^P = 3/2^-$ one. These two values are both consistent with the experimental mass of the $P_{cs}(4459)^0$, supporting its interpretation as the $\bar D^* \Xi_c$ molecular state of either $J^P=1/2^-$ or $3/2^-$. We also calculate the mass of the $J^P = 1/2^-$ $\bar D \Xi_c$ molecular state to be $4.29^{+0.13}_{-0.12}$~GeV.

To verify the hadronic molecular picture, we propose to further investigate the $P_{cs}(4459)^0$ state in future experiments to examine whether it can be separated into two states, and to search for the $J^P = 1/2^-$ $\bar D \Xi_c$ molecular state. If the hadronic molecular picture turns out to be correct, our understanding on the construction of the subatomic world would be significantly improved. Besides, these studies are helpful to improve our understanding on the non-perturbative behaviors of the strong interaction at the low energy region.

$\\$
{\it Hidden-charm pentaquark currents.}-----
We use the $\bar c$, $c$, $s$, $u$, and $d$ quarks to construct hidden-charm pentaquark interpolating currents with strangeness. To study $\bar D^{(*)} \Xi_c$ molecular states, we consider the following type of currents:
\begin{equation}
\eta(x)  = [\bar c_a(x) \Gamma_1 u_b(x)] ~ \Big[[d^T_c(x) \mathbb{C} \Gamma_2 s_d(x)] ~ \Gamma_3 c_e(x) \Big]   \, ,
\end{equation}
where $a \cdots e$ are color indices, $\Gamma_{1/2/3}$ are Dirac matrices, and $\mathbb{C} = i\gamma_2 \gamma_0$ is the charge-conjugation operator. The other type of currents:
\begin{equation}
\eta^\prime(x)  = [\bar c_a(x) \Gamma_1 d_b(x)] ~ \Big[[u^T_c(x) \mathbb{C} \Gamma_2 s_d(x)] ~ \Gamma_3 c_e(x) \Big]   \, ,
\end{equation}
can be similarly studied, but they just lead to the same QCD sum rule results as the $\eta(x)$ currents. Hence, the present study can not distinguish the isospin of $P_{cs}$ states.

The $\eta(x)$ currents can be constructed by combining charmed meson operators and charmed baryon fields. We need the charmed meson operators $J_{\mathcal{D}}$, which couple to the ground-state charmed mesons $\mathcal{D} = \bar D^0/\bar D^{*0}$:
\begin{eqnarray}
J_{\bar D^0} &=& \bar c_d \gamma_5 u_d \, ,
\\ \nonumber J_{\bar D^{*0}} &=& \bar c_d \gamma_\mu u_d \, .
\end{eqnarray}
We also need the charmed baryon field $J_{\mathcal{B}}$, which couples to the ground-state charmed baryon $\mathcal{B} = \Xi_c^0$:
\begin{eqnarray}
J_{\Xi_c^0} &=& \epsilon^{abc} [d_a^T \mathbb{C} \gamma_{5} s_b] c_c \, .
\label{eq:heavybaryon}
\end{eqnarray}

There can be altogether three $\bar D^{(*)} \Xi_c$ hadronic molecular states, that are $\bar D \Xi_c$ of $J^P = {1/2}^-$, $\bar D^* \Xi_c$ of $J^P = {1/2}^-$, and $\bar D^* \Xi_c$ of $J^P = {3/2}^-$. Their relevant interpolating currents are:
\begin{eqnarray}
\nonumber \eta_1 &=& [\delta^{ab} \bar c_a \gamma_5 u_b] ~ [\epsilon^{cde} d_c^T \mathbb{C} \gamma_5 s_d c_e]
\\ &=& \bar D^0 ~ \Xi_c^{0} \, ,
\\[1mm] \nonumber \eta_2 &=& [\delta^{ab} \bar c_a \gamma_\nu u_b] ~ \gamma^\nu \gamma_5 ~ [\epsilon^{cde} d_c^T \mathbb{C} \gamma_5 s_d c_e]
\\ &=& \bar D^{*0}_\nu ~ \gamma^\nu \gamma_5 ~ \Xi_c^{0} \, ,
\\[1mm] \nonumber \eta_3^\alpha &=& P_{3/2}^{\alpha\nu} ~ [\delta^{ab} \bar c_a \gamma_\nu u_b] ~ [\epsilon^{cde} d_c^T \mathbb{C} \gamma_5 s_d c_e]
\\ &=& P_{3/2}^{\alpha\nu} ~ \bar D^{*0}_\nu ~ \Xi_c^{0} \, .
\end{eqnarray}
In the above expressions, we have used $\mathcal{D}$ and $\mathcal{B}$ to denote the charmed meson operators $J_{\mathcal{D}}$ and the charmed baryon field $J_{\mathcal{B}}$; $P_{3/2}^{\mu\nu}$ is the spin-3/2 projection operator
\begin{equation}
P_{3/2}^{\mu\nu} = g^{\mu\nu} - {1 \over 4} \gamma^\mu\gamma^\nu \, .
\end{equation}

$\\$
{\it QCD sum rule studies.}-----
We use the method of QCD sum rules~\cite{Shifman:1978bx,Reinders:1984sr} to study $\bar D^{(*)} \Xi_c$ molecular states through the currents $\eta_{1,2}$ of $J^P = 1/2^-$ and $\eta^\alpha_{3}$ of $J^P = 3/2^-$. Taking the current $\eta_2$ as an example, we assume it couples to the $\bar D^* \Xi_c$ molecular state of $J^P = 1/2^-$ through
\begin{eqnarray}
\nonumber \langle 0 | \eta_2 | X; 1/2^- \rangle &=& f_X u (p) \, ,
\end{eqnarray}
where $u(p)$ is the Dirac spinor of this state, denoted as $X$ for simplicity. The two-point correlation function extracted from $\eta_2$ can be written as:
\begin{eqnarray}
\nonumber \Pi\left(q^2\right) &=& i \int d^4x e^{iq\cdot x} \langle 0 | T\left[\eta_{2}(x) \bar \eta_{2}(0)\right] | 0 \rangle
\\ \label{pi:spin12} &=& (q\!\!\!\slash~ + M_{X}) ~ \Pi_0\left(q^2\right) \, .
\end{eqnarray}

In QCD sum rule studies we calculate the two-point correlation function $\Pi_0\left(q^2\right)$ at both hadron and quark-gluon levels. At the hadron level, we use the dispersion relation to write it as
%
\begin{equation}
\Pi_0(q^2)={\frac{1}{\pi}}\int^\infty_{s_<}\frac{{\rm Im} \Pi_0(s)}{s-q^2-i\varepsilon}ds \, ,
\label{eq:disper}
\end{equation}
%
where $s_<$ is the physical threshold. We further define the imaginary part of the correlation function as the spectral density $\rho(s)$, which is usually evaluated at the hadron level by inserting intermediate hadron states $\sum_n|n\rangle\langle n|$:
%
\begin{eqnarray}
\rho_{\rm phen}(s) &\equiv& {\rm Im}\Pi_0(s)/\pi
\label{eq:rho}
\\ \nonumber &=& \sum_n\delta(s-M^2_n)\langle 0|\eta|n\rangle\langle n|{\eta^\dagger}|0\rangle
\\ \nonumber &=& f_X^2\delta(s-m_X^2)+ \mbox{continuum}.
\end{eqnarray}
%
In the last step we have adopted the usual parametrization of one-pole dominance for the ground state $X$ and a continuum contribution.

At the quark-gluon level we calculate $\Pi_0\left(q^2\right)$ using the method of operator product expansion (OPE), and derive its corresponding spectral density $\rho_{\rm OPE}(s)$. After performing the Borel transformation to Eq.~(\ref{eq:disper}) at both hadron and quark-gluon levels, we can approximate the continuum using the spectral density above a threshold value $s_0$ (quark-hadron duality), and obtain the sum rule equation
%
\begin{equation}
\Pi_0(s_0, M_B^2) \equiv f^2_X e^{-M_X^2/M_B^2} = \int^{s_0}_{s_<} e^{-s/M_B^2}\rho_{\rm OPE}(s)ds \, .
\label{eq:borel}
\end{equation}
%
It can be used to further calculate $M_X$ through
%
\begin{eqnarray}
M^2_X(s_0, M_B) &=& \frac{\int^{s_0}_{s_<} e^{-s/M_B^2}s\rho_{\rm OPE}(s)ds}{\int^{s_0}_{s_<} e^{-s/M_B^2}\rho_{\rm OPE}(s)ds} \, .
\label{eq:mass}
\end{eqnarray}
%

In the present study we calculate OPEs at the leading order of $\alpha_s$ and up to the $D({\rm imension}) = 10$ terms, including the perturbative term, the charm quark mass, the quark condensates $\langle \bar q q \rangle/\langle \bar s s \rangle$, the gluon condensate $\langle g_s^2 GG \rangle$, the quark-gluon mixed condensates $\langle g_s \bar q \sigma G q \rangle/\langle g_s \bar s \sigma G s \rangle$, and their combinations. We summarized the obtained spectral densities $\rho_{1\cdots3}(s)$ in Appendix~\ref{app:ope}, which are extracted from the currents $\eta_{1\cdots3}$, respectively. In the calculations we ignore the chirally suppressed terms with the up and down quark masses, and adopt the factorization assumption of vacuum saturation for higher dimensional condensates. We find that the $D=3$ quark condensates $\langle \bar q q \rangle/\langle \bar s s \rangle$ and the $D=5$ mixed condensates $\langle g_s \bar q \sigma G q \rangle/\langle g_s \bar s \sigma G s \rangle$ are multiplied by the charm quark mass, which are thus important power corrections.

$\\$
{\it Numerical analyses.}--------
We still use the current $\eta_2$ as an example to perform numerical analyses, where we use the following values for various QCD sum rule parameters~\cite{pdg,Yang:1993bp,Eidemuller:2000rc,Narison:2002pw,Gimenez:2005nt,Jamin:2002ev,Ioffe:2002be,Ovchinnikov:1988gk,Ellis:1996xc,colangelo}:
%
\begin{eqnarray}
\nonumber m_s &=& 96 ^{+8}_{-4} \mbox{ MeV} \, ,
\\ \nonumber m_c &=& 1.275 ^{+0.025}_{-0.035} \mbox{ GeV} \, ,
\\ \nonumber  \langle\bar qq \rangle &=& -(0.240 \pm 0.010)^3 \mbox{ GeV}^3 \, ,
\\ \langle\bar ss \rangle &=& (0.8\pm 0.1)\times \langle\bar qq \rangle \, ,
\label{condensates}
\\ \nonumber  \langle g_s^2GG\rangle &=& (0.48\pm 0.14) \mbox{ GeV}^4 \, ,
\\
\nonumber \langle g_s\bar q\sigma G q\rangle &=& - M_0^2\times\langle\bar qq\rangle \, ,
\\
\nonumber \langle g_s\bar s\sigma G s\rangle &=& - M_0^2\times\langle\bar ss\rangle \, ,
\\
\nonumber M_0^2 &=& (0.8 \pm 0.2) \mbox{ GeV}^2 \, .
\end{eqnarray}
Here the running mass in the $\overline{MS}$ scheme is used for the charm quark.

There are two free parameters in Eqs.~(\ref{eq:mass}): the Borel mass $M_B$ and the threshold value $s_0$. We use two criteria to constrain the Borel mass $M_B$ for a fixed $s_0$. The first criterion is used to insure the convergence of the OPE series. It is done by requiring the $D=10$ terms \big($m_c \langle \bar q q \rangle^3$ and $\langle g_s \bar q \sigma G q \rangle^2$\big) to be less than 20\%, which can determine the lower limit $M_B^{min}$:
%
\begin{equation}
\mbox{Convergence (CVG)} \equiv \left|\frac{ \Pi^{D=10}(\infty, M_B) }{ \Pi(\infty, M_B) }\right| \leq 20\% \, .
\label{eq:cvg}
\end{equation}
%
This criterion leads to $\left(M_B^{min}\right)^2 = 2.93$~GeV$^2$, when setting $s_0 = 25.8$~GeV$^2$.

The second criterion is used to insure the validity of one-pole parametrization. It is done by requiring the pole contribution to be larger than 50\%, which can determine the upper limit $M_B^{max}$:
%
\begin{equation}
\mbox{Pole Contribution (PC)} \equiv \frac{ \Pi(s_0, M_B) }{ \Pi(\infty, M_B) } \geq 50\% \, .
\label{eq:pole}
\end{equation}
%
This criterion leads to $\left(M_B^{max}\right)^2 = 3.07$~GeV$^2$, when setting $s_0 = 25.8$~GeV$^2$.

\begin{figure}[hbt]
\begin{center}
\scalebox{0.65}{\includegraphics{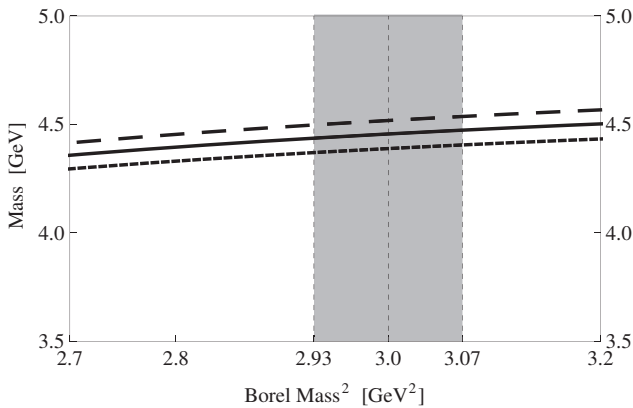}}
\caption{The variation of $M_X$ with respect to the Borel mass $M_B$, calculated using the current $\eta_2$. The short-dashed, solid, and long-dashed curves are obtained by setting $s_0 = 24.8$, $25.8$, and $26.8$ GeV$^2$, respectively.}
\label{fig:mass}
\end{center}
\end{figure}

Altogether we extract the working region of Borel mass to be $2.93$~GeV$^2< M_B^2 < 3.07$~GeV$^2$ for the current $\eta_2$ with the threshold value $s_0 = 25.8$~GeV$^2$. We show the variation of $M_{X}$ with respect to the Borel mass $M_B$ in Fig.~\ref{fig:mass} in a broader region $2.7$~GeV$^2\leq M_B^2 \leq 3.2$~GeV$^2$, and find it more stable inside the above Borel window.

Redoing the same procedures by changing $s_0$, we find that there are non-vanishing Borel windows as long as $s_0 \geq s_0^{min} = 24.8$~GeV$^2$. Accordingly, we choose the threshold value $s_0$ to be about 1.0~GeV larger with the uncertainty $\pm1.0$~GeV, {\it i.e.}, $24.8$~GeV$^2\leq s_0\leq 26.8$~GeV$^2$. Altogether, our working regions for the current $\eta_2$ are determined to be $24.8$~GeV$^2\leq s_0\leq 26.8$~GeV$^2$ and $2.93$~GeV$^2< M_B^2 < 3.07$~GeV$^2$, where the mass is extracted to be:
\begin{equation}
M_{\bar D^* \Xi_c; 1/2^-} = 4.46^{+0.16}_{-0.14} \mbox{ GeV} \, .
\end{equation}
Here the central value corresponds to $M_B^2=3.00$ GeV$^2$ and $s_0 = 25.8$ GeV$^2$. Its uncertainty comes from the Borel mass $M_B$, the threshold value $s_0$, the charm quark mass $m_c$, and various QCD sum rule parameters listed in Eqs.~(\ref{condensates}). This mass value is consistent with the experimental mass of the $P_{cs}(4459)^0$, supporting its interpretation as the $\bar D^* \Xi_c$ molecular state of $J^P=1/2^-$.

Similarly, we use the currents $\eta_{1}$ and $\eta^\alpha_{3}$ to perform numerical analyses, and extract masses of the $J^P=1/2^-$ $\bar D \Xi_c$ molecular state and the $J^P=3/2^-$ $\bar D^* \Xi_c$ molecular state to be:
\begin{eqnarray}
M_{\bar D \Xi_c; 1/2^-}   &=& 4.29^{+0.13}_{-0.12} \mbox{ GeV} \, ,
\\
M_{\bar D^* \Xi_c; 3/2^-} &=& 4.47^{+0.19}_{-0.15} \mbox{ GeV} \, .
\end{eqnarray}
Hence, our results also support the interpretation of the $P_{cs}(4459)^0$ as the $\bar D^* \Xi_c$ molecular state of $J^P=3/2^-$. We summarize all the above results in Table~\ref{tab:mass}.

Generally speaking, understanding the nature of exotic hadrons is a complicated topic, since different structures with the same quantum numbers can contribute to the same state, and different structures may have similar masses. However, these different structures may lead to different decay processes. Therefore, to determine whether the $P_{cs}(4459)^0$ is the $\bar D^* \Xi_c$ molecular state of $J^P=1/2^-$ or the one of $J^P=3/2^-$ in our framework, we shall further calculate its width in our future study, to be compared with its experimental value $\Gamma_{P_{cs}(4459)^0} = 17.3 \pm  6.5^{+8.0}_{-5.7}$~MeV~\cite{lhcb}.

\begin{table*}[hpt]
\begin{center}
\renewcommand{\arraystretch}{1.5}
\caption{Masses of the $\bar D^{(*)} \Xi_c$ hadronic molecular states, extracted from the currents $\eta^{(\alpha)}_{1\cdots3}$.}
\begin{tabular}{c | c | c | c | c | c | c}
\hline\hline
\multirow{2}{*}{\,Currents\,} & \multirow{2}{*}{\,Configuration\,} & \multirow{2}{*}{~$s_0^{min}~[{\rm GeV}^2]$~} & \multicolumn{2}{c|}{Working Regions} & \multirow{2}{*}{~Pole~[\%]~} & \multirow{2}{*}{~Mass~[GeV]~}
\\ \cline{4-5} & & & ~~$s_0~[{\rm GeV}^2]$~~ & ~~$M_B^2~[{\rm GeV}^2]$~~ &
\\ \hline \hline
$\eta_1$                  & $|\bar D \Xi_c; 1/2^- \rangle$               & $22.3$ & $23.3\pm1.0$ & $2.61$--$2.77$ & $50$--$56$ & $4.29^{+0.13}_{-0.12}$
\\
$\eta_2$                  & $|\bar D^* \Xi_c; 1/2^- \rangle$             & $24.8$ & $25.8\pm1.0$ & $2.93$--$3.07$ & $50$--$55$ & $4.46^{+0.16}_{-0.14}$
\\
$\eta_3^\alpha$           & $|\bar D^* \Xi_c; 3/2^- \rangle$             & $24.4$ & $25.4\pm1.0$ & $2.81$--$2.95$ & $50$--$55$ & $4.47^{+0.19}_{-0.15}$
\\ \hline\hline
\end{tabular}
\label{tab:mass}
\end{center}
\end{table*}

$\\$
{\it Summary and Discussions.}-----
Very recently, the LHCb Collaboration reported the evidence of a hidden-charm pentaquark state with strangeness, $P_{cs}(4459)^0$, in the $J/\psi \Lambda$ invariant mass spectrum of the $\Xi_b^- \to J/\psi K^- \Lambda$ decays~\cite{lhcb}. This state contains at least five quarks $\bar c c s qq$ ($q=u/d$), with one of them the $strange$ quark. This LHCb experiment indicates that there probably exist many more exotic hadrons with strangeness to be discovered in the near future, so a new hadron spectrum is waiting to be constructed.

The $P_{cs}(4459)^0$ is about 19~MeV below the $\bar D^{*0} \Xi_c^0$ threshold, so it is natural to interpret it as the $\bar D^{*} \Xi_c$ molecular state. Accordingly, in this letter we use the method of QCD sum rules to study the $\bar D^{*} \Xi_c$ molecular states of $J^P=1/2^-$ and $3/2^-$, and at the same time we also study the $\bar D \Xi_c$ molecular state of $J^P=1/2^-$. We evaluate their masses to be
\begin{eqnarray*}
M_{\bar D \Xi_c; 1/2^-}   &=& 4.29^{+0.13}_{-0.12} \mbox{ GeV} \, ,
\\
M_{\bar D^* \Xi_c; 1/2^-} &=& 4.46^{+0.16}_{-0.14} \mbox{ GeV} \, ,
\\
M_{\bar D^* \Xi_c; 3/2^-} &=& 4.47^{+0.19}_{-0.15} \mbox{ GeV} \, .
\end{eqnarray*}
Hence, our QCD sum rule results support the interpretation of the $P_{cs}(4459)^0$ as the $\bar D^* \Xi_c$ molecular state of either $J^P=1/2^-$ or $3/2^-$.

%
\begin{figure*}[hbt]
\begin{center}
\subfigure[~$\Xi_b^- \to \bar D^{(*)0} \Xi_c^{0} K^-$]{\includegraphics[width=0.32\textwidth]{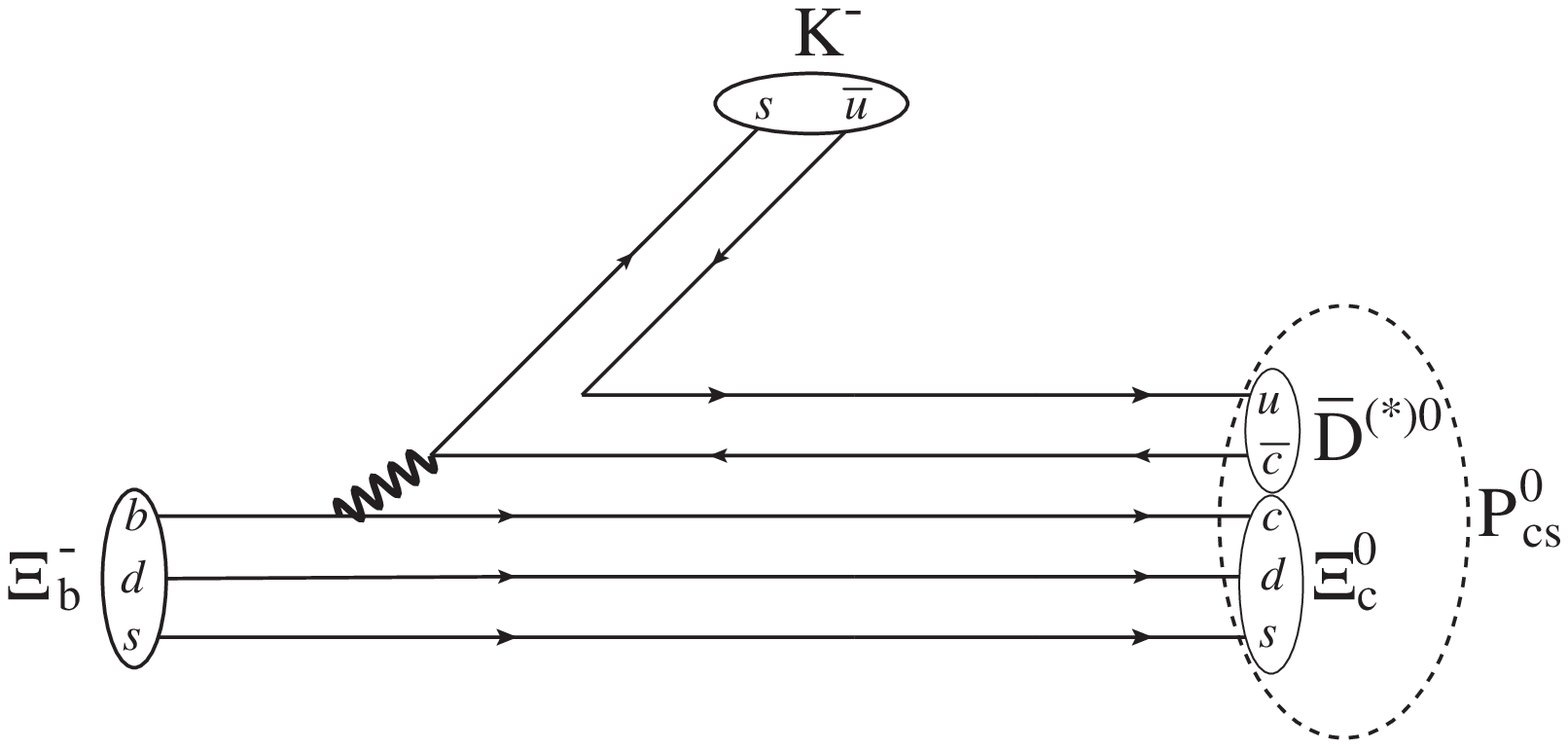}}
~
\subfigure[~$\Lambda_b^0 \to \bar D^{(*)0} \Lambda_c^{+} K^-$]{\includegraphics[width=0.32\textwidth]{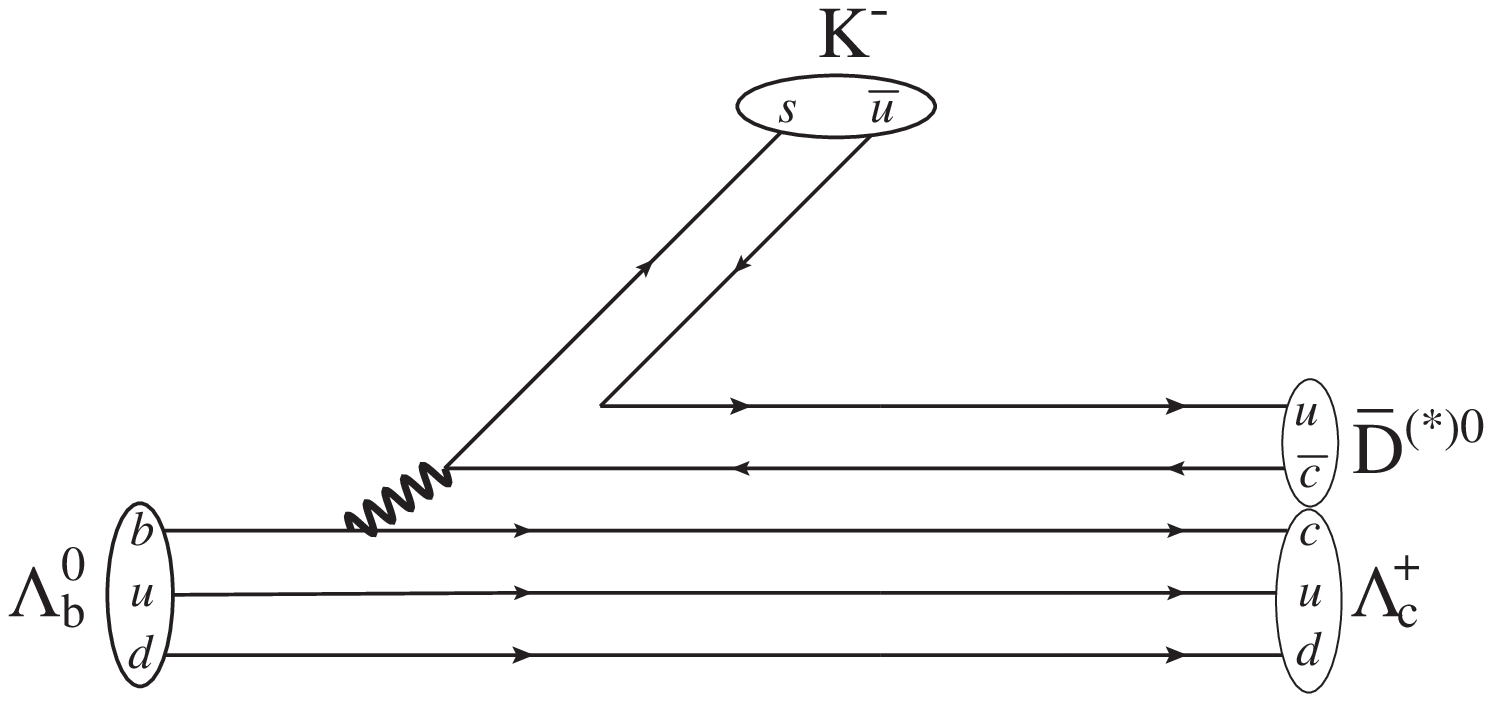}}
~
\subfigure[~$\Lambda_b^0 \to \bar D^{(*)0} \Sigma_c^{(*)+} K^-$]{\includegraphics[width=0.32\textwidth]{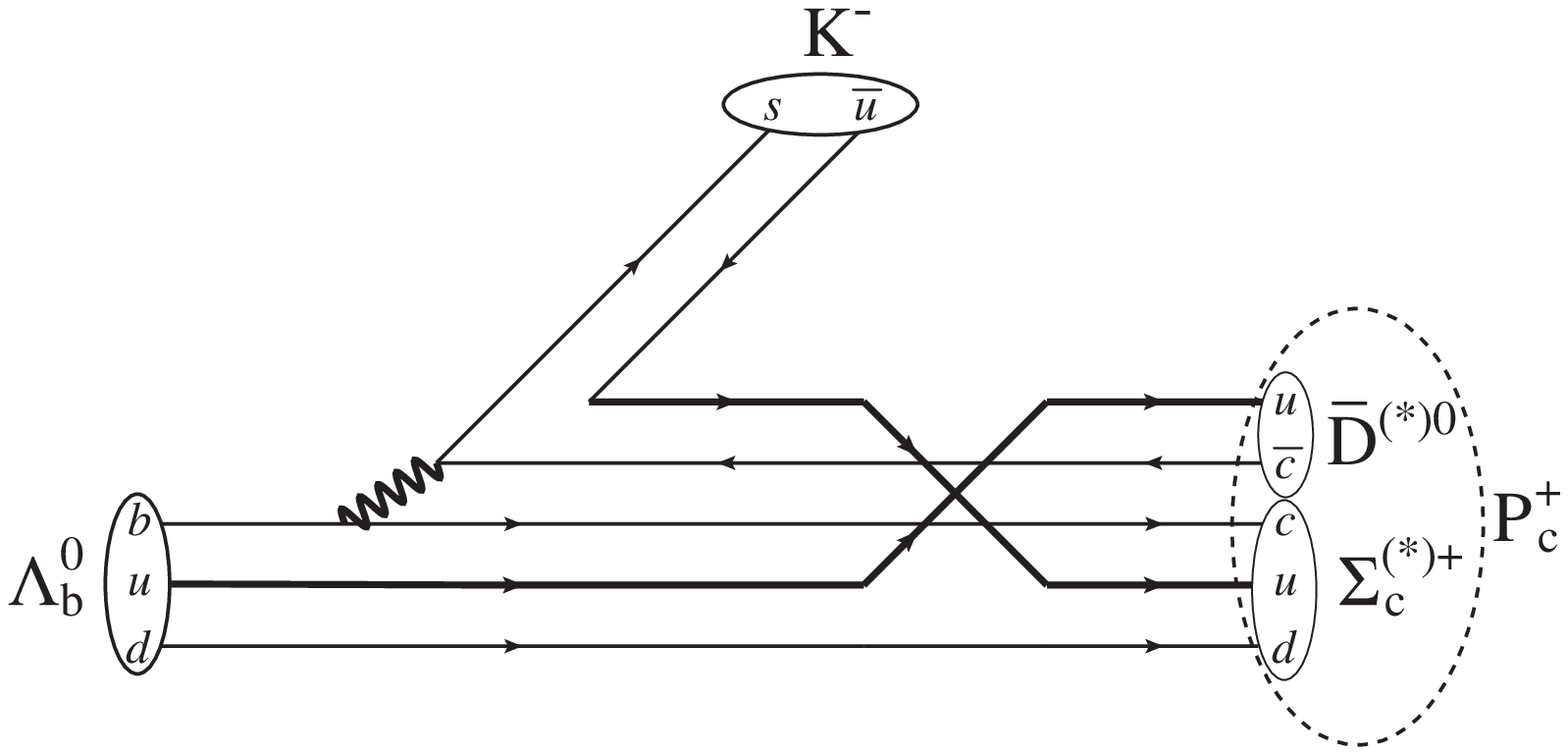}}
\caption{Possible production mechanisms of the $P_c^+$ and $P_{cs}^0$ states in $\Lambda_b^0/\Xi_b^-$ decays.}
\label{fig:production}
\end{center}
\end{figure*}
%

Within the hadronic molecular picture, the three LHCb experiments observing $P_c$ and $P_{cs}$ states~\cite{lhcb,Aaij:2015tga,Aaij:2019vzc} can be well understood as a whole:
\begin{itemize}

\item The transition of $\Xi_b^-\to J/\psi K^-\Lambda$ is dominated by the Cabibbo-favored weak decay of $b\to c + \bar{c}s$ via the $V$-$A$ current. This leads to an intuitive expectation that the $s$ and $d$ pair in $\Xi_b^-$ may exist as a spectator, so that their total spin $S=0$ is conserved. As a consequence, if this $sd$ pair is to be combined with a $c$ quark to form a charmed baryon, it will favor the $\Xi_c$ instead of the $\Xi_c^\prime$ and $\Xi_c^*$, of which the total spin of the $sd$ pair is $S = 1$. Accordingly, the $P_{cs}(4459)^0$ possibly as the $\bar D^* \Xi_c$ molecular state was observed in this channel by LHCb~\cite{lhcb}, other than those possibly existing $\bar D^{(*)} \Xi_c^\prime$ and $\bar D^{(*)} \Xi_c^*$ molecular states. We illustrate this process in Fig.~\ref{fig:production}(a).

\item The transition of $\Lambda_b^0\to J/\psi K^- p$ can be similarly analysed, which may favor the $\Lambda_c$ instead of the $\Sigma_c$ and $\Sigma_c^*$. However, in this case the $\Lambda_c$ is probably not bounded with the charmed mesons due to the lack of $\pi$ exchanges. It is only because of the significantly larger data sample (680k for the $\Lambda_b^0\to J/\psi K^- p$ decays and only 4k for the $\Xi_b^-\to J/\psi K^-\Lambda$ decays~\cite{lhcb}), that the $P_c(4312)$, $P_c(4440)$, and $P_c(4457)$ possibly as the $\bar D \Sigma_c$ and $\bar D^{*} \Sigma_c$ molecular states were observed in this channel by LHCb~\cite{Aaij:2015tga,Aaij:2019vzc}. We illustrate the relevant two processes in Fig.~\ref{fig:production}(b,c).

\end{itemize}
Therefore, the three LHCb experiments observing $P_c$ and $P_{cs}$ states~\cite{lhcb,Aaij:2015tga,Aaij:2019vzc} strongly support the hadronic molecular picture and the existence of ``hadronic molecules''.

To further verify the above hadronic molecular picture, we propose to investigate the $P_{cs}(4459)^0$ in future experiments to examine whether it can be separated into two states. We also propose to search for the $\bar D \Xi_c$ molecular state of $J^P=1/2^-$, whose mass is predicted to be $4.29^{+0.13}_{-0.12}$~GeV. If the hadronic molecular picture turns out to be correct, our understanding on the construction of the subatomic world be significantly improved, and our understanding on the non-perturbative behaviors of the strong interaction at the low energy region would also be significantly improved.

\section*{Acknowledgments}

This project is supported by
the National Natural Science Foundation of China under Grants No.~11722540, No.~11975165, and No.~12075019,
the China National Funds for Distinguished Young Scientists under Grant No. 11825503,
the National Program for Support of Top-notch Young Professionals,
and
the Fundamental Research Funds for the Central Universities.

\appendix

\begin{widetext}
\section{Spectral densities}
\label{app:ope}

In this appendix we list the spectral densities $\rho_{1\cdots3}(s)$ extracted for the currents $\eta_{1\cdots3}$. In the following expressions, $\mathcal{F}(s) = \FF(s)$, $\mathcal{H}(s) = \HH(s)$, and the integration limits are $\alpha_{min}=\frac{1-\sqrt{1-4m_c^2/s}}{2}$, $\alpha_{max}=\frac{1+\sqrt{1-4m_c^2/s}}{2}$, $\beta_{min}=\frac{\alpha m_c^2}{\alpha s-m_c^2}$, and $\beta_{max}=1-\alpha$. In the calculations we take into account both the $m_s$ and $m_s^2$ terms, but here we list only the $m_s$ terms.

The spectral density $\rho_{1}(s)$ extracted for the current $\eta_{1}$ is
\begin{eqnarray}
\rho_{1}(s) &=& \rho^{pert}_{1}(s) + \rho^{\qq}_{1}(s) + \rho^{\GGa}_{1}(s)+ \rho^{\qGqa}_{1}(s) + \rho^{\qq^2}_{1}(s)  + \rho^{\qq\qGqa}_{1}(s)+ \rho^{\qGqa^2}_{1}(s) + \rho^{\qq^3}_{1}(s) \, ,
\end{eqnarray}
where
\begin{eqnarray}
\nonumber \rho^{pert}_{1}(s) &=& \dab \Bigg\{ \mathcal{F}(s)^5 \times  \frac{(1 - \alpha - \beta)^3}{163840 \pi ^8 \alpha ^5 \beta ^4} \Bigg\} \, ,
\non
\rho^{\qq}_{1}(s) &=& \dab \Bigg\{ \mathcal{F}(s)^3 \times  \frac{(1 - \alpha - \beta) ( (\alpha +\beta -1) \beta m_c \qq  + \beta  m_s (\sss - 2 \qq))}{2048 \pi ^6 \alpha ^3 \beta ^3} \Bigg\} \, ,
\non
\rho^{\GGa}_{1}(s) &=& {\GGb } \dab\Bigg\{ m_c^2 \mathcal{F}(s)^2    \times \frac{(1 - \alpha - \beta)^3 \left(\alpha ^3+\beta ^3\right)}{196608 \pi ^8 \alpha ^5 \beta ^4}
\non && ~~~~~~ + \mathcal{F}(s)^3 \times    \frac{(\alpha +\beta -1) \left(3 \alpha ^3-\alpha ^2 (\beta +3)-2 \alpha  (\beta -1) \beta -(\beta -1)^2 \beta \right)}{196608 \pi ^8 \alpha ^5 \beta ^3}   \Bigg\} \, ,
\non
\rho^{\qGqa}_{1}(s) &=& \dab \Bigg\{ \mathcal{F}(s)^2 \times \frac{-3 \left(\alpha ^2+\alpha  (3 \beta -2)+2 \beta ^2-3 \beta +1\right){m_c\qGqb }  + 3 \beta ^2 {m_s\qGqb } }{4096 \pi ^6 \alpha ^2 \beta ^3} \Bigg\} \, ,
\non
\rho^{\qq^2}_{1}(s)&=& \dab \Bigg\{ \frac{- \mathcal{F}(s)^2 \times \qq \sss + m_c m_s \mathcal{F}(s) \times \alpha (2 \qq^2 - \qq \sss)}{256 \pi ^4 \alpha ^2 \beta }   \Bigg\} \, ,
\non
\rho^{\qq\qGqa}_{1}(s)&=& \int^{\alpha_{max}}_{\alpha_{min}}d\alpha  \Bigg\{ \int^{\beta_{max}}_{\beta_{min}}d\beta \Bigg\{ \frac{m_c m_s \qGqb (2 \qq -  \sss )}{512 \pi ^4 \beta }            \Bigg\}
\non && ~~~~~~ +  \mathcal{H}(s) \times      \frac{ \qq \sGsb + \sss \qGqb}{512 \pi ^4 \alpha }  -\frac{ m_c m_s \qGqb (4 \qq - \sss)}{1024 \pi ^4}                                                                                                                     \Bigg\} \, ,
\non
\rho^{\qGqa^2}_{1}(s)&=& \int^{\alpha_{max}}_{\alpha_{min}}d\alpha \Bigg\{          \frac{(\alpha -1) \qGqb \sGsb }{1024 \pi ^4}                                 \Bigg\}
+ \int^{1}_{0}d\alpha \Bigg\{ \delta\left(s - {m_c^2 \over \alpha(1-\alpha)}\right)
\non && ~~~~~~ \times  \frac{\qGqb \left(m_c^3 m_s \qGqb/M_B^2 + (1-\alpha) m_c^2 \sGsb - \alpha  (\alpha +1) m_c m_s \qGqb \right)}{2048 \pi ^4 (\alpha -1) \alpha }      \Bigg\} \, ,
\non
\rho^{\qq^3}_{1}(s)&=& {m_c\qq^2 \sss } \int^{\alpha_{max}}_{\alpha_{min}}d\alpha \Bigg\{       \frac{1}{96 \pi ^2}                            \Bigg\} \, .
\end{eqnarray}

The spectral density $\rho_{2}(s)$ extracted for the current $\eta_{2}$ is
\begin{eqnarray}
\rho_{2}(s) &=& \rho^{pert}_{2}(s) + \rho^{\qq}_{2}(s) + \rho^{\GGa}_{2}(s)+ \rho^{\qGqa}_{2}(s) + \rho^{\qq^2}_{2}(s)  + \rho^{\qq\qGqa}_{2}(s)+ \rho^{\qGqa^2}_{2}(s) + \rho^{\qq^3}_{2}(s) \, ,
\end{eqnarray}
where
\begin{eqnarray}
\nonumber \rho^{pert}_{2}(s) &=& \dab \Bigg\{ \mathcal{F}(s)^5 \times \frac{(1 - \alpha - \beta)^3}{81920 \pi ^8 \alpha ^5 \beta ^4} \Bigg\} \, ,
\non
\rho^{\qq}_{2}(s) &=& \dab \Bigg\{ \mathcal{F}(s)^3 \times \frac{ (1 - \alpha - \beta) (2 (\alpha +\beta -1) m_c \qq + \beta m_s (\sss - 2 \qq))}{1024 \pi ^6 \alpha ^3 \beta ^3} \Bigg\} \, ,
\non
\rho^{\GGa}_{2}(s) &=& {\GGb } \dab\Bigg\{ m_c^2 \mathcal{F}(s)^2    \times \frac{ (1 - \alpha - \beta)^3 \left(\alpha ^3+\beta ^3\right)}{98304 \pi ^8 \alpha ^5 \beta ^4}
\non && ~~~~~~ + \mathcal{F}(s)^3 \times  \frac{ (1 - \alpha - \beta) \left(3 \alpha ^3+\alpha ^2 (7 \beta -3)+2 \alpha  (\beta -1) \beta +(\beta -1)^2 \beta \right)}{98304 \pi ^8 \alpha ^5 \beta ^3}  \Bigg\} \, ,
\non
\rho^{\qGqa}_{2}(s) &=& \dab \Bigg\{ \mathcal{F}(s)^2 \times \frac{- 3 \qGqb (2 (\alpha +\beta -1) m_c - \beta m_s)}{2048 \pi ^6 \alpha ^2 \beta ^2} \Bigg\} \, ,
\non
\rho^{\qq^2}_{2}(s)&=& \dab \Bigg\{ \frac{ -  \mathcal{F}(s)^2 \times \qq \sss - m_c m_s \mathcal{F}(s) \times 2 \alpha ( \sss - 2 \qq)}{128 \pi ^4 \alpha ^2 \beta } \Bigg\} \, ,
\non
\rho^{\qq\qGqa}_{2}(s)&=& \int^{\alpha_{max}}_{\alpha_{min}}d\alpha  \Bigg\{  \mathcal{H}(s) \times   \frac{\qq \sGsb + \sss \qGqb}{256 \pi ^4 \alpha }
-\frac{ m_c m_s \qGqb (4 \qq - \sss)}{256 \pi ^4}               \Bigg\} \, ,
\non
\rho^{\qGqa^2}_{2}(s)&=& \int^{\alpha_{max}}_{\alpha_{min}}d\alpha \Bigg\{     \frac{(\alpha -1) \qGqb \sGsb}{512 \pi ^4}                   \Bigg\}
+ \int^{1}_{0}d\alpha \Bigg\{ \delta\left(s - {m_c^2 \over \alpha(1-\alpha)}\right)
\non && ~~~~~~ \times  \frac{ \qGqb \left(2 m_c^3 m_s \qGqb / M_B^2 + ( 1-\alpha) m_c^2 \sGsb - 2 (\alpha -1) \alpha m_s \qGqb \right)}{1024 \pi ^4 (\alpha -1) \alpha }   \Bigg\} \, ,
\non
\rho^{\qq^3}_{2}(s)&=& {m_c\qq^2 \sss } \int^{\alpha_{max}}_{\alpha_{min}}d\alpha \Bigg\{    \frac{1}{24 \pi ^2}                \Bigg\} \, .
\end{eqnarray}

The spectral density $\rho_{3}(s)$ extracted for the current $\eta_{3}^\alpha$ is
\begin{eqnarray}
\rho_{3}(s) &=& \rho^{pert}_{3}(s) + \rho^{\qq}_{3}(s) + \rho^{\GGa}_{3}(s)+ \rho^{\qGqa}_{3}(s) + \rho^{\qq^2}_{3}(s)  + \rho^{\qq\qGqa}_{3}(s)+ \rho^{\qGqa^2}_{3}(s) + \rho^{\qq^3}_{3}(s) \, ,
\end{eqnarray}
where
\begin{eqnarray}
\nonumber \rho^{pert}_{3}(s) &=& \dab \Bigg\{ \mathcal{F}(s)^5 \times \frac{(1 - \alpha - \beta)^3 (\alpha +\beta +4)}{1310720 \pi ^8 \alpha ^5 \beta ^4} \Bigg\} \, ,
\non
\rho^{\qq}_{3}(s) &=& \dab \Bigg\{ \mathcal{F}(s)^3 \times \frac{(\alpha +\beta -1) (6  (1 - \alpha - \beta) m_c \qq + \beta (2 \alpha +2 \beta +3) m_s (2 \qq - \sss))}{16384 \pi ^6 \alpha ^3 \beta ^3} \Bigg\} \, ,
\non
\rho^{\GGa}_{3}(s) &=& {\GGb } \dab\Bigg\{ m_c^2 \mathcal{F}(s)^2    \times \frac{(1 - \alpha - \beta)^3 (\alpha +\beta +4) \left(\alpha ^3+\beta ^3\right)}{1572864 \pi ^8 \alpha ^5 \beta ^4}
\non && ~~~~~~ + \mathcal{F}(s)^3 \times \Bigg( \frac{ 4 \alpha ^5-3 \alpha ^4 (3 \beta +11)-6 \alpha ^3 \left(6 \beta ^2+13 \beta -9\right)-\alpha ^2 \left(32 \beta ^3+51 \beta ^2-108 \beta +25\right)}{4718592 \pi ^8 \alpha ^5 \beta ^3}
\non && ~~~~~~ ~~~~~~~~~~~~~~~~~  \frac{-3 \alpha  (\beta -1)^2 \beta  (4 \beta +11)-3 (\beta -1)^3 \beta  (\beta +4)}{4718592 \pi ^8 \alpha ^5 \beta ^3} \Bigg) \Bigg\} \, ,
\non
\rho^{\qGqa}_{3}(s) &=& \dab \Bigg\{ \mathcal{F}(s)^2 \times \frac{3 \qGqb (-6 (\alpha +\beta -1) m_c + \beta (4 \alpha +4 \beta +1) m_s)}{32768 \pi ^6 \alpha ^2 \beta ^2} \Bigg\} \, ,
\non
\rho^{\qq^2}_{3}(s)&=& \dab \Bigg\{ \frac{ - \mathcal{F}(s)^2 \times(4 \alpha +4 \beta +1) \qq \sss - m_c m_s \mathcal{F}(s) \times 6 \alpha (\sss - 2 \qq)}{2048 \pi ^4 \alpha ^2 \beta } \Bigg\} \, ,
\non
\rho^{\qq\qGqa}_{3}(s)&=& \int^{\alpha_{max}}_{\alpha_{min}}d\alpha  \Bigg\{ \int^{\beta_{max}}_{\beta_{min}}d\beta \Bigg\{\mathcal{F}(s) \times \frac{- \qq \sGsb - \sss \qGqb}{1024 \pi ^4 \alpha }      \Bigg\}
\non && ~~~~~~ +  \mathcal{H}(s) \times    \frac{5\qq \sGsb + 5 \sss \qGqb}{4096 \pi ^4 \alpha }  -\frac{3 m_c m_s \qGqb (4 \qq -  \sss )}{4096 \pi ^4}                                                                                                                   \Bigg\} \, ,
\non
\rho^{\qGqa^2}_{3}(s)&=& \int^{\alpha_{max}}_{\alpha_{min}}d\alpha \Bigg\{    \frac{3 (\alpha -1) \qGqb \sGsb}{8192 \pi ^4}          \Bigg\}
+ \int^{1}_{0}d\alpha \Bigg\{ \delta\left(s - {m_c^2 \over \alpha(1-\alpha)}\right)
\non && ~~~~~~ \times \frac{\qGqb \left(6 m_c^3 m_s \qGqb/M_B^2 - 5 (\alpha -1) m_c^2 \sGsb - 6 (\alpha -1) \alpha m_c m_s \qGqb \right)}{16384 \pi ^4 (\alpha -1) \alpha }  \Bigg\} \, ,
\non
\rho^{\qq^3}_{3}(s)&=& {m_c\qq^2 \sss } \int^{\alpha_{max}}_{\alpha_{min}}d\alpha \Bigg\{   \frac{1}{128 \pi ^2}            \Bigg\} \, .
\end{eqnarray}

\end{widetext}

\end{document}